\documentclass[12pt]{article}

\usepackage{times}
\usepackage{graphicx}
\usepackage[superscript,biblabel]{cite}
\usepackage{fancyhdr}
\usepackage{placeins}
\usepackage{floatflt}
\usepackage{epstopdf}
\usepackage{hyperref}
\usepackage[normalem]{ulem}
\usepackage{xcolor}
\usepackage{gensymb}
\begin{document}

\title{Autonomous Detection of Methane Emissions in Multispectral Satellite Data Using Deep Learning}
\author
{Bertrand Rouet-Leduc$^{1,2\dag}$, Thomas Kerdreux$^{1}$,\\
Alexandre Tuel$^{1}$, Claudia Hulbert$^{1\ast}$ \\
\normalsize{$^{1}$Geolabe, Los Alamos, New Mexico, USA}\\
\normalsize{$^{2}$Kyoto University, DPRI, Japan}\\
\normalsize{$^\ast$To whom correspondence should be addressed;}\\ 
\normalsize{E-mail: claudiah@geolabe.com}
}

\date{}
\maketitle

\begin{abstract}

Methane is one of the most potent greenhouse gases, and its short atmospheric half-life makes it a prime target to rapidly curb global warming. However, current methane emission monitoring techniques primarily rely on approximate emission factors or self-reporting, which have been shown to often dramatically underestimate emissions. Although initially designed to monitor surface properties, satellite multispectral data has recently emerged as a powerful method to analyze atmospheric content. However, the spectral resolution of multispectral instruments is poor, and methane measurements are typically very noisy. Methane data products are also sensitive to absorption by the surface and other atmospheric gases (water vapor in particular) and therefore provide noisy maps of potential methane plumes, that typically require extensive human analysis. Here, we show that the image recognition capabilities of deep learning methods can be leveraged to automatize the detection of methane leaks in Sentinel-2 satellite multispectral data, with dramatically reduced false positive rates compared with state-of-the-art multispectral methane data products, and without the need for a priori knowledge of potential leak sites. Our proposed approach paves the way for the automated, high-definition and high-frequency monitoring of point-source methane emissions across the world.

\end{abstract}

\newpage

\section{Introduction}

Methane is the second highest contributor to human-driven climate change after carbon dioxyde (CO$_2$). It is a potent greenhouse gas with a warming potential considerably higher than that of CO$_2$ (84 times higher over a 20-year horizon). Though emitted in much smaller amounts than CO$_2$, methane is thus estimated to be responsible for about a third of the warming to date\cite{IPCC2021}. Yet, methane has a half-life of only 10 years in the atmosphere, much shorter than that of CO$_2$. Reducing methane emissions would therefore bring substantial climate mitigation benefits in the short-to-medium term. 
Halving anthropogenic methane emissions by 2050 could avoid 0.25°C of warming by mid-century and 0.5°C by 2100\cite{Ocko2021} -- significant numbers, given that the Earth has already warmed by 1.1°C.

Recognizing the urgency of this challenge, governments have recently set out to develop far-reaching strategies to curb greenhouse gas emissions in general, and methane emissions in particular\cite{EPA2021,EU2021}. These are largely targeted towards industry, particularly the oil and gas sector which is estimated to account for 25-30\% of anthropogenic methane emissions\cite{saunois2020global}.\\ 

A cornerstone for measuring progress towards emission targets, ensuring compliance with new regulations, and prioritizing among remedial actions, is to be able to precisely detect and quantify the emissions in question. However, methods to precisely characterize greenhouse gas emissions at scale are currently lacking. Most countries worldwide do not require emitters to report data. Existing emission inventories are not based on a direct monitoring of all possible emission sources, but instead on normative models, \textit{i.e.} models that apply an empirical average emission coefficient (emission factor) to estimated production volumes from various industries. As a result, there currently are major gaps between emission estimates based on local atmospheric measurements and official reports, and we do not have a clear picture of emissions\cite{IEAglobal2022}. When they exist, self-reported data are known to underestimate emissions, often substantially. For instance, a recent study\cite{thorpe2020methane} showed that methane emissions at a dozen underground gas storage (UGS) sites in California, estimated through airborne spectral imagery, were five times higher than reported. Similarly, methane emissions from the New Mexico Permian Basin have recently been found to be 6.5 times larger than reported\cite{chen2022}. \\

Ground and airborne monitoring approaches (e.g., in-situ sensors, foot patrols, drones, planes) are costly and intrinsically limited to the monitoring of localized methane emissions. Consequently, they cannot be used to detect and quantify methane at a large scale, to identify unknown sources of emissions, or to create coherent regional, national or global methane inventories. By contrast, satellite-based monitoring of methane provides the opportunity to monitor emissions at scale, which explains why spectral satellite imagery has attracted much attention over the last few years for quantifying emissions. Emissions are identified in data produced by spectral satellites: as a gas emission occurs, it modifies the reflectance measured by spectral sensors over the area where the emission is contained. This change in reflectance occurs unevenly in the spectral domain, as gases preferably absorb light at specific wavelengths. The comparison between such changes and theoretical changes induced by the known spectra of methane can be used to characterize methane emissions in spectral images. \\

Hyperspectral and multispectral satellite constellations such as Sentinel-5P, Sentinel-2, Prisma, GoSat, WorldView-3 or the commercial GHGsat constellation, have been used successfully to detect methane emissions from space \cite{varon2018quantifying, irakulis2021satellites, cusworth2021multisatellite, lauvaux2022global, maasakkers2022}, demonstrating their potential for monitoring emissions at scale. However, given the already large (several hundreds of TB) and upcoming volume of data produced by satellite constellations spanning large regions or even the entire Earth at high spatial and temporal resolution, new algorithms are needed to analyze spectral satellite data in order to detect and quantify methane emissions in an computationally efficient way. \\

An ideal methane monitoring system at scale should combine satellite data with a high spatial resolution and frequent revisit time along with automated and computationally efficient algorithms, in order to seamlessly detect, pinpoint and quantify localized methane sources across large areas with minimal human intervention. In what follows, we propose a new method for identifying emissions in multispectral data. Because of its low spectral resolution, few studies rely on multispectral satellites to detect methane emissions \cite{varon2021high, ehret2022global}. However, the high spatial resolution (20 m) and good repeat pass (a few days) of multispectral satellites make them an ideal choice compared to other alternatives. We show that using deep learning to analyze these images allows to considerably improve detection thresholds in multispectral images from the Sentinel-2 and Landsat 8 constellations. This technology could ultimately be deployed as a fully automatic monitoring system allowing to identify emissions and responsible actors, anywhere on Earth, with a repeat pass of only 5 days that could be easily improved by including data from other constellations.\\

Most existing methods for methane detection rely on methane total column concentration data, computed using spectral images from hyperspectral satellites. This approach has the disadvantage of adding a processing step that can sometimes be computationally lengthy, and imposes to rely on satellite constellations that typically have drawbacks in terms of either spatial coverage or spatial resolution. There are today no existing machine learning methods that detect methane in satellite multispectral imagery. The very few machine learning approaches developed for the detection of greenhouse gases in spectral airbone \cite{Jongaramrungruang2022} or satellite\cite{finch2022} data all take as input methane concentration maps obtained from hyperspectral imagery. By contrast, we propose a methodology where emissions are directly identified from multispectral signals, without relying on intermediate processing that may hinder the computational efficiency of the approach.\\

In a first demonstration of our methodology, we analyze known methane leaks in Algeria and at the Aliso Canyon underground gas storage site, and show that a simple deep learning method dramatically outperforms state-of-the-art methane products based on multispectral band ratios. Our method consistently detects the leaks during their lifetime of a few months, with very few false positive detections, paving the way for fully automatic methane sources detection at high frequency and global scale.

\section{Results}
\subsection{Deep learning for methane point source detection in multispectral data}

\subsubsection*{Fully convolutional auto-encoder approach}

Our goal is to detect methane point sources from noisy multispectral data. Multispectral instruments such as Sentinel-2's MSI were designed to analyze the Earth's surface, and their capability to detect methane sources was a lucky by-product of the missions. Sentinel-2 offers high spatial resolution of 10-60 m per pixel and a high return frequency of 2-6 days at global scale\cite{Li2017}. However, the poor spectral resolution in discrete bands results in noisy methane data products: band 12 of the MSI instrument intersects the absorption spectrum of methane, but also that of water and CO$_2$. The absorption spectra of water and CO$_2$ also intersect the other bands of the instrument, which is not the case for methane. Methane estimates therefore rely on extracting salient reductions in measured ground reflectance that appear in band B12 but not in the other bands. The resulting products are nevertheless very noisy and still require human interpretation and a priori knowledge of possible methane leak sources \cite{varon2021high}.\\

Here we essentially task a deep learning model with learning the spectral footprint of methane plumes in multispectral images that is devoid of false positives and tentatively is only sensitive to methane, and not to other gases or to ground reflectance features. Our model must be able to distinguish the spatial and temporal statistical differences between methane absorption signals and noise from the absorption of the rest of the atmosphere and noise from surface composition and texture.\\
In order to separate signal from atmospheric and surface noise, we developed the deep learning architecture shown in Fig. \ref{fig:fullCNN_schematic}. This architecture consists in 5 purely convolutional layers. The first 3 layers of the model are tasked with encoding signals by learning appropriate band ratios, and the last 2 layers are tasked with decoding the signal at the original spatial resolution.

\subsubsection*{Training on synthetic data}

Deep learning models require large amounts of data to be trained, and no (or extremely limited) ground truth exists for methane detection in multispectral data. We therefore rely on synthetic data to train our deep learning models, in the spirit of previous applications we have developed for extracting small signals from radar satellite data \cite{rouetleduc2021} as well as to detect small elasto-gravity signals\cite{licciardi2022}. To this end we model the effect of synthetic methane plumes as reduced reflectance in about 700 real Sentinel-2 scenes over a variety of regions and continents. The Sentinel-2 data consists of L1C Top-of-Atmosphere (TOA) reflectance computed from data measured by the spectrometers onboard Sentinel-2A and 2B. These images are made of $13$ different spectral bands with a square pixel resolution varying from 10 to 60 meters. Band $12$ (and marginally band $11$) intersects with the methane absorption spectrum. The input bands are upsampled or downsampled to the 20 meter resolution of bands $11$ and $12$. We take care to avoid the northern region of Algeria where a real methane leak took place in 2019 and on which the performance of our method is demonstrated below. 

\begin{figure*}[!ht]
  \begin{center}
    \includegraphics[width=0.95\linewidth]{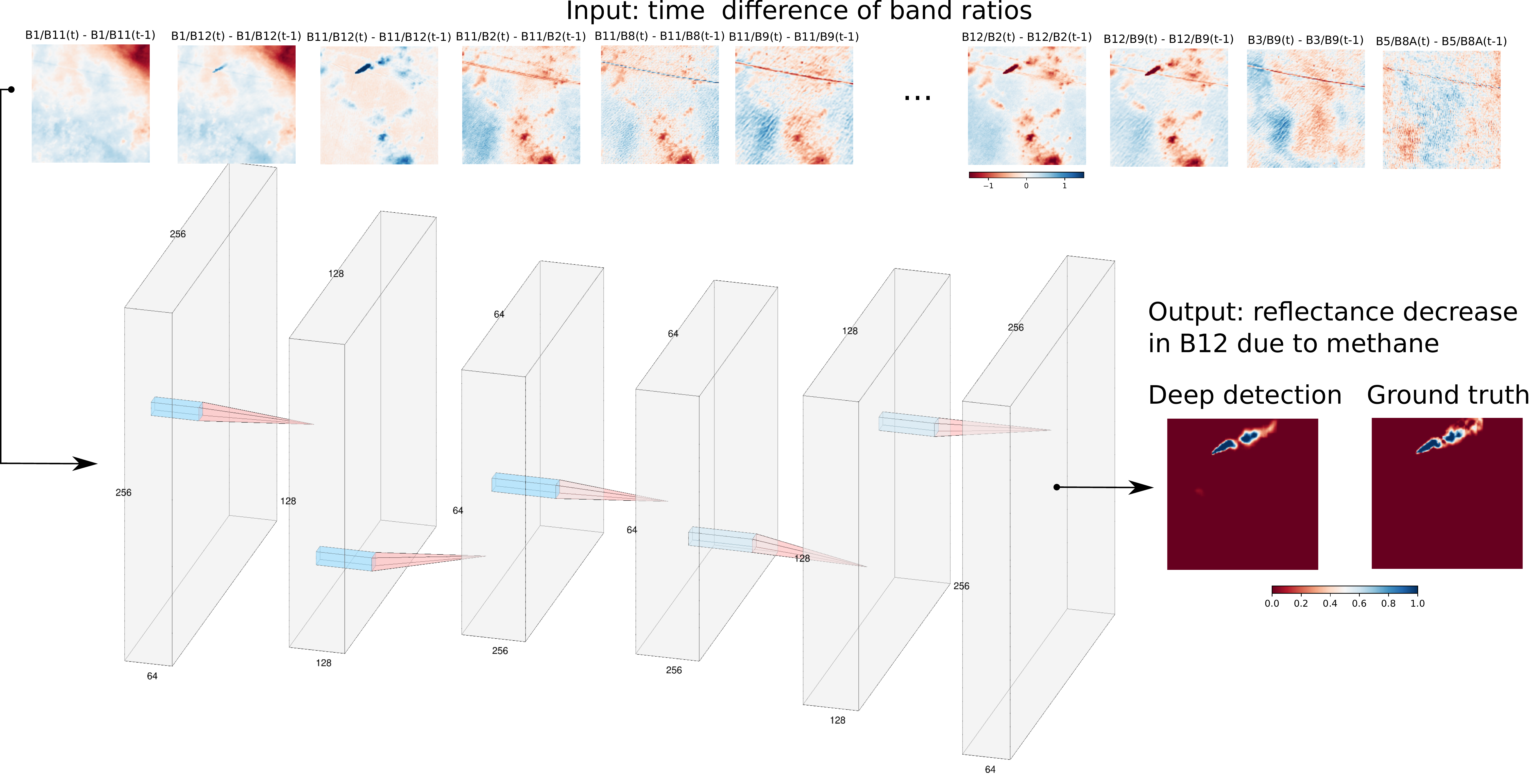}
    \caption{\textbf{Deep auto-encoder trained to detect methane plumes on synthetic data}. Schematic of our deep learning auto-encoder model. Top row (left to right): a sequence of synthetic multispectral Sentinel-2 data on which the model is trained, where synthetic methane plumes are embedded inside the images using the Beer-Lambert law. Second row: the architecture of our model. Our model is purely convolutional with progressive pooling on the spatial dimension during the encoding. The last layers of the model are tasked with decoding the mask of reduced reflectance due to the methane plume, here compared with ground truth to the right.}
     \label{fig:fullCNN_schematic}
  \end{center}
\end{figure*}
\FloatBarrier

To build a training dataset of synthetic multispectral satellite images of methane leaks, we assume the leaks stem from point sources. Instead of physically realistic WRF-LES simulation schemes \cite{Yamaguchi2012}, we opted for custom Gaussian plumes to efficiently generate millions of different patterns, with the goal of creating a diverse training dataset without attempting to simulate detailed physical processes. We generated thousands of Gaussian plumes with various emission rates and wind velocities, to which we added auto-correlated atmospheric noise to emulate atmospheric turbulence. The plumes were then embedded into the Sentinel-2 scenes using the Beer-Lambert law\cite{ehret2022global}.\\

The input fed to the network is a stack of normalized time differences of all possible band ratios between bands B1, B2, B3, B4, B5, B8, B8A, B9, B11 and B12 (45 combinations), taken over two neighbouring time steps (here conceptually noted $t-1$ and $t$, though they need not correspond to consecutive satellites passes). An example is shown on Fig. \ref{fig:fullCNN_schematic}, where the third image input in the top row is the normalized time difference of band ratio B11 over B12. The task of the neural network is to find the leak in the input images, defined as the ensemble of pixels where absorbance in band 12 (that overlaps with the methane absorption spectrum) has decreased by at least 1\% at time $t$.

All 691,201 trainable parameters are optimized to minimize binary cross-entropy loss on this training data, with the Adam variation of stochastic gradient descent. The model is trained to find synthetic plumes embedded in about 1 million real Sentinel 2 samples drawn from about 500 Sentinel-2 tiles from various regions chosen for their diversity and sampled throughout the year 2022: Afghanistan, Argentina, Canada, Chile, China, Egypt, Ethiopia, France, India and Iran (with each tile containing about 1800 unique 128$\times$128 samples).
About 130,000 validation samples are drawn from 100 tiles from Japan and Kenya, and the model we retain is the one with the best performance on this validation dataset. The model is finally evaluated in Fig. \ref{fig:MBMP_vs_CNN} on performance on bout 130,000 test samples drawn from 100 tiles from Mali and Mexico.

\subsection{Performance on synthetic test data}\label{perf_synthetic}

Before applying our model to real methane leaks, we first assess its performance on test data consisting of synthetic plumes (from a different set of synthetic plumes than that used in training) embedded in samples from Sentinel-2 tiles that are distinct from those used for training.
The deep neural network is trained using the binary cross-entropy of finding the pixels where band 12 reflectance is reduced by more than 5\% due to the methane from a leak. In order to assess the model using metrics that have a more intuitive understanding, we assess here the performance of the model in terms of the error on segmenting the methane plume (i.e. the classification error of pixels containing methane above 5\% of reflectance reduction in band 12). \\

\begin{figure*}[!ht]
  \begin{center}
    \includegraphics[width=0.95\linewidth]{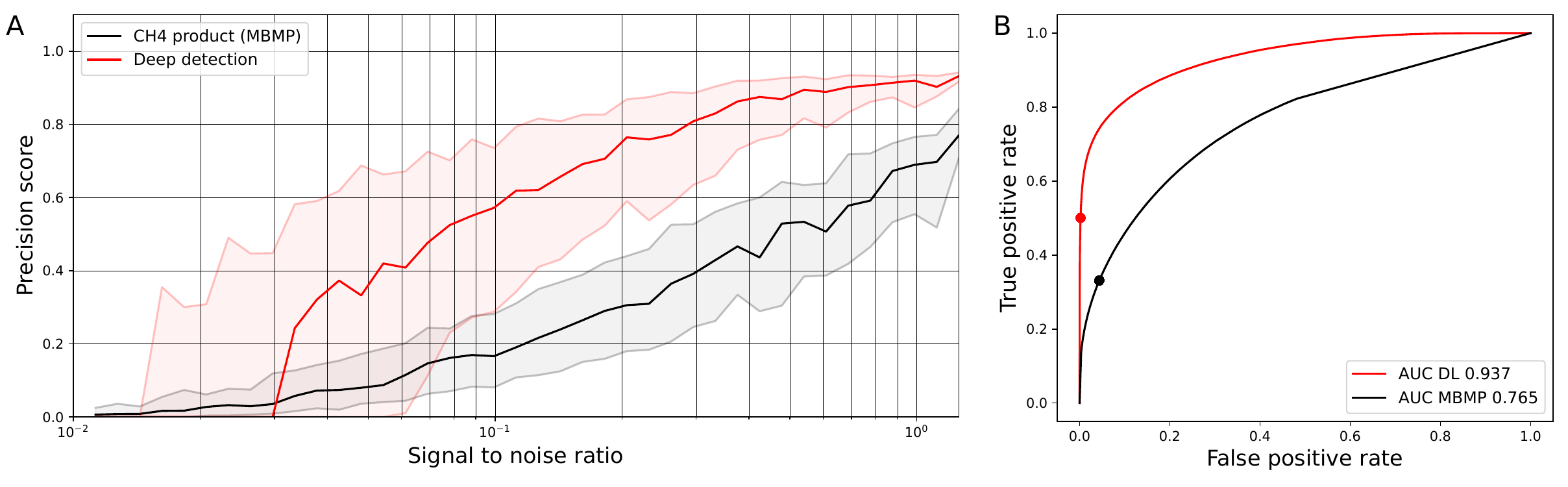}
    
    \caption{\textbf{Performance of our deep learning approach versus the multi-band-multi-pass (MBMP) approach.} \textbf{A.} Performance in terms of precision score (ratio of true positives to all positives for the pixel-wise presence of the plume (defined as 5\% or more reflectance reduction in band 12) of our deep learning model (in red), compared with the MBMP method \cite{varon2021high} (in black), as a function of signal-to-noise-ratio (SNR).   \textbf{B.} Area under the receiver-operator curve (ROC-AUC) that shows the ratio of true positives to false positives for various classifying thresholds and for various signal to noise ratio conditions, with the thresholds of 0.5 used in \textbf{A} shown as dots Note that for a threshold of 0.5 the model detects most leaks in the synthetic data, while making less than 1\% false positives, compared with 20\% false positives when using a threshold on the MBMP method that detects 50\% of the plumes. The absence of false positives determines the level of automation of a method. The data used for this evaluation is a test set: the deep learning model has not been trained on this dataset, that comes from a different region than the region used for training, and the testing set synthetic plumes are different from the training set plumes. }
     \label{fig:MBMP_vs_CNN}
  \end{center}
\end{figure*}

The left plot of Fig. \ref{fig:MBMP_vs_CNN} shows the classification performance of our deep learning model, in terms of precision score (ratio of true positives to true and false positives) as a function of signal-to-noise ratio (SNR). The SNR is defined as the ratio of the mean reflectance reduction in band 12 due to the embedded synthetic plume (within the mask of the plume) to the standard deviation of reflectance in band 12 before embedding (in the whole image). We compare our model to the multi-band multi-pass (MBMP) method\cite{varon2021high}, with a plume detection threshold chosen such that the MBMP method has perfect precision score for high SNR. We find that that the MBMP method suffers from very high false positive rates for SNR $<$ 1 while our deep learning performs satisfactorily down to SNR $\approx$ 0.1. At SNRs below 1, it becomes difficult for the eye to notice the reduction in reflectance due to the plumes, and we argue that our model reaches super-human performance. 

We argue that, as shown by the ROC-AUC shown in right plot of Fig. \ref{fig:MBMP_vs_CNN}, using a conservative decision threshold a deep learning classifier can be made that has very few false positives (e.g. less than 1\%) and still detects most of the methane leaks in SNR conditions between 10 to 100\%, while the MBMP approach suffers from high false positive rates at SNR lower than 1, even for conservative detection thresholds. A model that is able to detect most leaks with practically no false positives can then be used in an automated fashion.

\FloatBarrier
\FloatBarrier

The performance of our deep detection of synthetic methane leaks in Sentinel 2 data gives an estimate of the detection threshold of our model in terms of signal to noise ratio, and demonstrates of detection threshold around 0.1 SNR using deep learning, versus about 1 using a MBMP method. There are caveats to this performance comparison: the model is trained in a region of Algeria and tested in a different region of Algeria, but will not have the same performance for other regions with dramatically different backgrounds. We will show in the next sections that the model still generalizes to different regions and is able to detect methane leaks in California for example. For the best performance the model would need to be retrained for different regions or on a dataset including a variety of regions. A second caveat is that our deep learning model and the MBMP method shown here only use two time steps, and using more timesteps will improve the performance of both, and the difference in performance may not be as dramatic.

\subsection{Application to real methane leaks}

\subsubsection{Unlit flare in the Hassi Messaoud oil field (Algeria) in 2019}

\begin{figure*}[!ht]
  \begin{center}
    \includegraphics[width=0.95\linewidth]{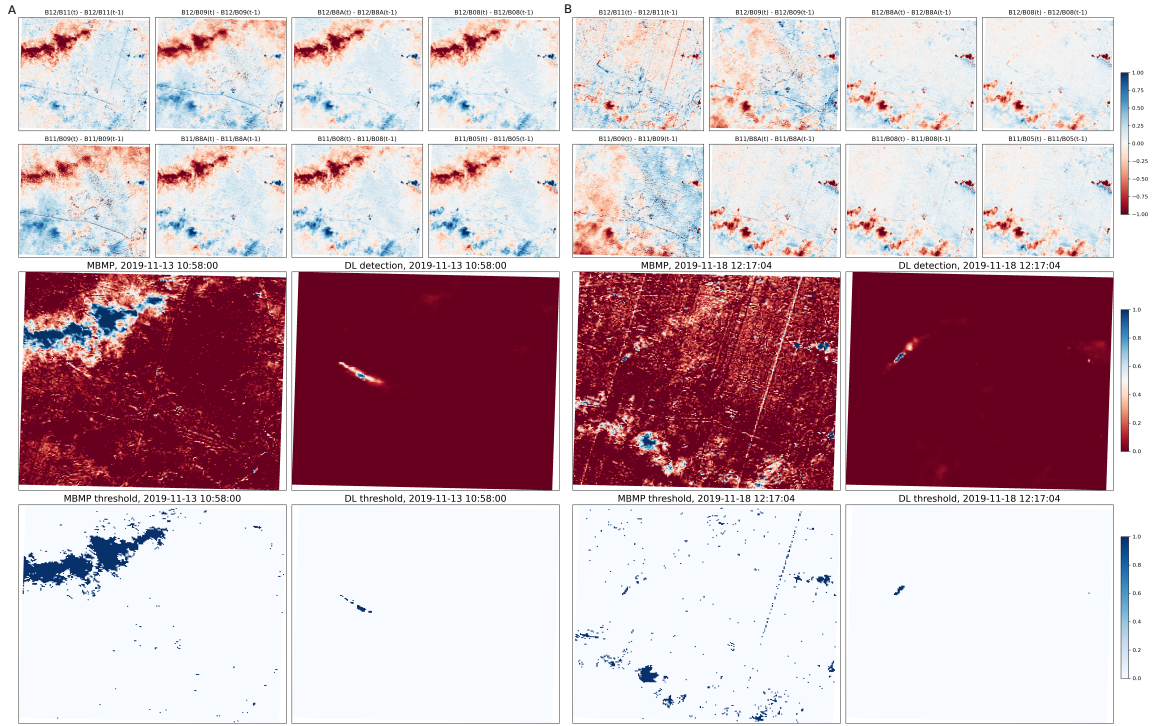}
    \caption{\textbf{Detecting a known methane leak in Algeria: deep detection versus MBMP under varying atmospheric conditions}. Here we feed our neural network with the time difference of Sentinel-2 TOA reflectance band ratios measured over a known leak in northern Algeria between \textbf{A)} November 13, 2019 and November 8, 2019 , and \textbf{B)} November 18, 2019 and November 13, 2019. In both A and B, the first two rows show a subset of the input data (time difference of band ratios) fed to our deep neural network. We include the band 12/band 11 ratio in which the methane signature is the most pronounced. The middle left plots in A and B show the plume detection based on a threshold applied to the time difference of the band 12/band 11 ratio, which is similar to the multi-band multi-pass (MBMP) method\cite{varon2021high}. The middle right plots in A and B show the output of our deep detector, with sharp signals on the known leak and very little (presumably) false positives elsewhere. The lower plots show a classification of the plume location based on applying the same 0.5 threshold to both methods, as in Fig. \ref{fig:MBMP_vs_CNN}.}
     \label{fig:Algeria}
  \end{center}
\end{figure*}
\FloatBarrier

Our purely convolutional auto-encoder trained to detect simulated plumes embedded in real Sentinel-2 data from southern Algeria can be applied on arbitrarily-sized data, enabling the straightforward analysis of known historical leaks or on-the-fly data. Here, we test our model on a known months-long leak from the Hassi Messaoud oil field in northern Algeria. The corresponding data was not used during the training of the model. Our model consistently detects the leak, with very few detections besides the known methane leak (which would presumably be false positives) without the need for further human analysis or a priori knowledge of possible leak locations.

For comparison, the time difference of band ratios using bands B11 and B12 is akin to the state of the art for methane leak detection\cite{varon2021high, zhang2022}, but is extremely noisy because many phenomena (clouds, surface albedo changes, etc.) create plume-like absorption features in bands B11 and B12. This approach tends to perform well in ideal atmospheric conditions (see Fig. \ref{fig:Algeria}A), but struggles in the presence of noise, especially when clouds or CO2 plumes are present (see Fig. \ref{fig:Algeria}B). Note that Figs. \ref{fig:Algeria}A and \ref{fig:Algeria}B refer to the same unlit flare at the exact same location (but on two dates with different wind directions).

These results illustrate the ability of deep learning models to discover combinations of band ratios and spatial and temporal features that are typical of methane leaks, dramatically reducing false positives and thereby outperforming standard approaches.

\subsubsection{Leak rate detection threshold in real Sentinel 2 data}

\begin{figure*}[!ht]
  \begin{center}
    \includegraphics[width=0.8\linewidth]{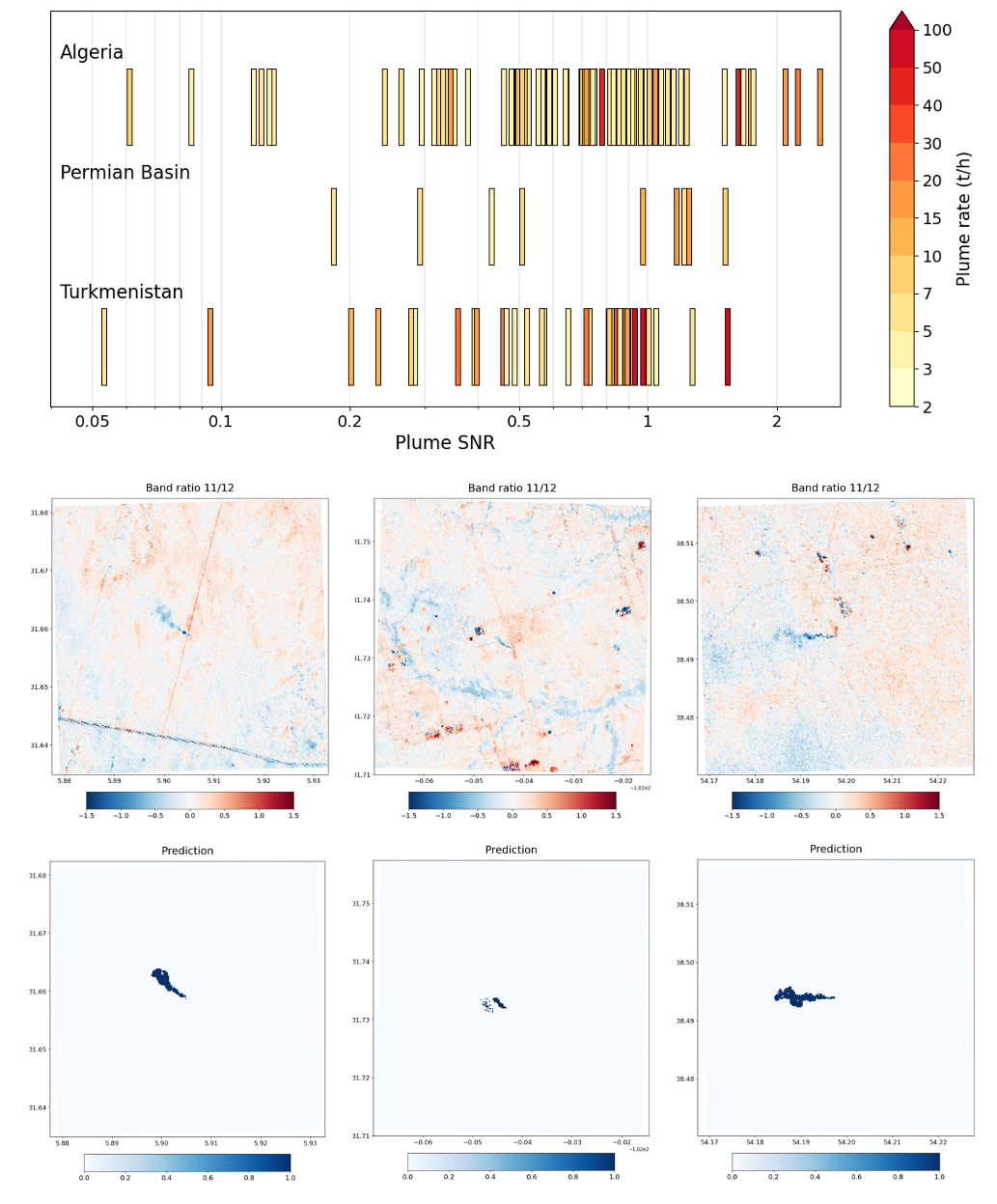}
    \caption{\textbf{Performance of our deep learning algorithm on real methane leaks}. The top panel shows the distribution of detected methane plumes in terms of SNR (x-axis) and emission rate (shading, in t/h) for the three regions of interest: Algeria, the Permian Basin and Turkmenistan. The bottom panels show three random examples of B11/B12 ratios and the corresponding model prediction, for Algeria (left), the Permian Basin (center) and Turkmenistan (right).}
     \label{fig:plume_rates}
  \end{center}
\end{figure*}

 After estimating the methane leak detection performance of our model as a function of SNR for embedded synthetic methane plumes (section \ref{perf_synthetic}), we now detect real methane plumes known from the literature, whose leak rate we estimate using the Integrated Mass Enhancement method (IME, see Methods) in three regions of interest: the Hassi Messaoud oil field in Algeria (already shown in Fig. \ref{fig:Algeria}), the Permian Basin in the U.S. and the Korpezhe oil and gas field in Turkmenistan.\\
 For this exercise, we are able to detect automatically the Hassi Messaoud oil field leak for 72 of the 91 cloud-free Sentinel-2 acquisitions during which the leak is present (between October 2019 and August 2020), as well as the Permian Basin leak for 4 of the 9 corresponding cloud-free Sentinel-2 acquisitions (between July and September 2020), and the leak in Turkmenistan for 45 of the 50 corresponding cloud-free Sentinel-2 acquisitions. In Fig. \ref{fig:plume_rates}, we show the known leak rate of the real methane plumes our deep detector is able to detect, and the corresponding SNR estimate, showing that our model appears to reliably and automatically detect plumes in varied conditions for leak rates above 2-3 tons per hour.

\FloatBarrier


\section{Discussion}

Satellite multispectral instruments were not initially designed to detect greenhouse gas emissions, and methane products derived from these instruments are therefore extremely noisy. For this reason, expert processing and analysis is required to interpret methane data products, and analyses are focused on areas of known potential leaks.

Furthermore, since the onset of the Sentinel-2 mission, the amount of available multispectral data has grown at a pace that is already challenging the ability of the community to process and analyze it. Automatic, autonomous spectral data interpretation methods are poised to become essential, if just to leverage the increasing spatial and temporal resolution of the data.
Our approach is applicable with little adjustment to data from other multispectral satellite constellations (\textit{e.g.} Landsat 8), and can also be modified for application to constellations with higher spectral or spatial resolution.

The initial application of our method on multispectral Sentinel-2 time series enables the direct observation of a methane leak, refining previous detections, autonomously and without prior knowledge. We expect our method to provide the ability to systematically and automatically observe methane leaks at a global scale.

\section*{Methods}

\subsection*{Deep learning model architecture}

The deep learning architecture designed for methane detection from multispectral images is shown on Figure \ref{fig:fullCNN_schematic}. Our model consists in 5 purely convolutional layers and as a result can be applied to arbitrarily-sized inputs. The first two convolutional layers are followed by max-pooling layers, while the last two layers are transposed convolutional layers. All the layers of the network are followed by ReLu activation functions. At each layer a set of filters is learned and except for the first layer that takes an input with as many channels as there are band ratios, all the other convolutional layers of the network learn a set of convolutional filters to be applied in the spatial domain. The band ratios with bands B1, B11, B12, B2, B3, B4, B5, B8, B8A, B9 form 45 possibilities, that are as many channels of the input to our network. The training samples in which we embed the synthetics are of size 256 x 256 pixels. After the first layer the data is turned into 256 x 256 images with 64 channels, and to 128 x 128 images with 128 channels after the second layer, and to 64 x 64 images with 256 channels after the third layer. After the fourth layer the data goes back to 128 x 128 images with 128 channels, and the model finally outputs a single channel 256 x 256 image of the reduced reflectance in band 12 due to methane. Our model is thus a fully convolutional deep denoiser that considers spectral signal as noise except for that of reduced reflectance in band 12 that is due to methane, considered the only signal here for our purposes.

Each layer has a number of trainable parameters given by $n_{\mathrm{kernel}}\times n_{\mathrm{input}}\times n_{\mathrm{output}}+ n_{\mathrm{output}}$, with $n_{\mathrm{kernel}}$ the convolutional kernel size (product of its shape in all dimensions), $n_{\mathrm{input}}$ the number of input channels to the layer, and $n_{\mathrm{output}}$ the number of output channels of the layer. This gives $3\times3\times45\times64+64$ trainable parameters for the first encoding layer, $3\times3\times64\times128+128$ trainable parameters for the second encoding layer, $3\times3\times256\times256+256$ trainable parameters for the third encoding layer, $3\times3\times256\times128+128$ trainable parameters for the first decoding layer, $3\times3\times128\times1+1$ trainable parameters for the second decoding layer, and $3\times3\times64\times1+1$ trainable parameters for the second decoding layer, for a total of $691,201$ trainable parameters. Our model was implemented and trained on GPUs using the Pytorch library.

\subsection*{Evaluation metrics}

The normalized root mean squared error NRMSE we used to assess the performance of the models in Fig. \ref{fig:MBMP_vs_CNN} is defined as the Euclidian norm of the residual error normalized by the Euclidian norm of the ground truth: $\mathrm{NRMSE(y,\hat{y})= \frac{\|y-\hat{y}\|}{\|y\|}}$, with $y$ the ground truth of reduced reflectance in band 12 due to methane and $\hat{y}$ its estimation.\\
The segmentation error estimated in \ref{fig:MBMP_vs_CNN} is assessed with the precision score $=\frac{\mathrm{TP}}{\mathrm{TP+FP}}$ (ratio of true positives TP to true positives and false positives FP) of classifiers built from our deep denoiser as well as from the MBMP method using thresholds to turn the continuous outputs into classes (methane detection and no methane detection). 
In training the synthetic data is normalized such that output reduced band 12 reflectance due to methane is bound between 0 and 1. The corresponding classes are made using a threshold of 0.5 (above which we consider there is methane).
The classifier built from our deep learning model is correspondingly when our model's output is above the same threshold of 0.5. Building a classifier from the MBMP method is slightly less straightforward as it is not bound. Thus, we first clip the MBMP output between 0 and 1 and then apply the same threshold of 0.5 to consider a detection. Treated in this manner both methods converge towards the same perfect precision of 1 for high signal to noise ratios. Precision is defined as the ratio of true positives to true and false positives.

\subsection*{Integrated Mass Enhancement (IME) method}
 The Integrated Mass Enhancement (IME) method estimates the rate of a detected methane plume based on the total plume methane mass detected downwind of the source\cite{Frankenberg2016, varon2018quantifying}. With $\Delta\Omega$ the excess methane field (in kg m$^{-2}$) and $\mathcal{P}$ the plume mask, the plume methane mass is given by:
 \begin{equation}
     IME = \sum_{i\in \mathcal{P}} \Delta\Omega(i) A(i)
 \end{equation}
 \noindent where $A(i)$ is the area of grid cell $i$ (in m$^2$). The plume rate $Q$ is then empirically related to $IME$ by way of an effective wind speed $U_{eff}$ (in m s$^{-1}$), and an effective plume size $L$ (in m) through:
 \begin{equation}
     Q = \frac{U_{eff} IME}{L}
 \end{equation}
 Varon et al. (2018)\cite{varon2018quantifying} found that $U_{eff}$ could be derived from the local 10-meter wind speed $U_{10}$ through $U_{eff} = 0.9\log(U_{10})+0.6$. We obtain 10-meter wind speed from the hourly GEOS-FP global reanalysis\cite{Molod2015}. Finally, we retrieve $\Delta\Omega$ from the Sentinel-2 B11/B12 band ratio values by comparing their values with band ratios simulated by the libRadtran radiative model\cite{Emde2016} under the same solar/satellite geometric conditions\cite{varon2021high}.

\subsection*{Data availability}
All the Sentinel 2 data used here is freely available from the European Space Agency on various repositories, such as PEPS used for this study (\url{https://peps.cnes.fr}). The Hassi Messaoud oil field leak we analysed is located at 31.6585\degree N/5.9053\degree E, and we downloaded Sentinel-2 scenes for the corresponding area from October 1, 2019 to August 31, 2020. The methane leak in Turkmenistan's Korpezhe oil and gas field is located at 38.4939\degree N/54.1977\degree E, and we downloaded the corresponding Sentinel-2 data from July 1, 2017 to December 31, 2020. Both leaks were analysed by Varon et al. (2018)\cite{varon2021high}. Finally, we analysed the methane leak in the Permian Basin described by Ehret et al. (2022) \cite{ehret2022global} located at 31.7335\degree N/102.0421\degree W from July 1, 2020 to September 30, 2020.\\
The GEOS-FP reanalysis data is available from \url{https://fluid.nccs.nasa.gov/weather/}.

\subsection*{Acknowledgements}
This work was funded by the U.S. Department of Energy under the SBIR grant DE-SC0022398.

\bibliographystyle{naturemag}
\bibliography{utils/biblio}

\appendix

\section{The Aliso Canyon reservoir failure in 2015}

\begin{figure*}[!ht]
   \begin{center}
     \includegraphics[width=0.7\linewidth]{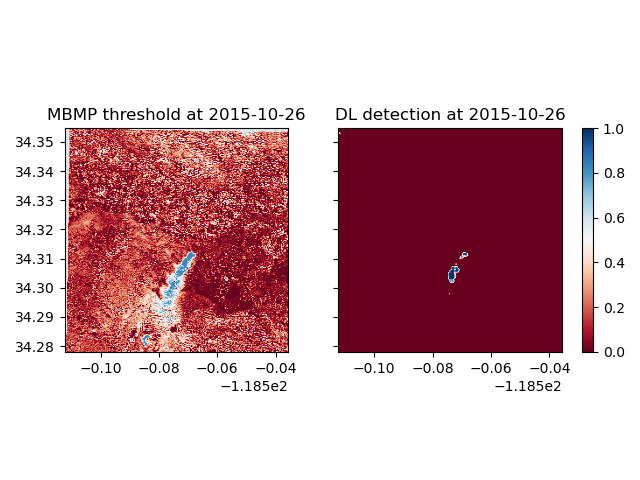}
     \caption{\textbf{Detecting a known methane leak in California: the Aliso reservoir 2015 accident}. As in Fig. \ref{fig:Algeria}, we feed our trained neural network the time difference of LandSat-8 band ratios measured over the Aliso Canyon reservoir failure in 2015. The output of our deep detector (lower right) shows a sharp signal on the known leak and very little (presumably) false positives elsewhere although the spectrometer differs from that of Sentinel-2.  However, note that the Aliso Canyon leak emitted several tens of tons of methane per hour.}
      \label{fig:Aliso_LandSat}
   \end{center}
 \end{figure*}

Here we illustrate the ability of our model to generalize to data from different regions and from different sensors on which it has been trained. In Fig. \ref{fig:Aliso_LandSat}, we show the application of our model to LandSat-8 data of the Aliso Canyon reservoir leak in late 2015.

The model does correctly identify the methane plume, in spite of a different background noise from which it has been trained on and in spite of the difference between the Sentinel-2 spectral band and the LandSat-8 bands.
For comparison, we show the results from the same threshold on the MBMP method as in Fig. \ref{fig:Algeria}, which also identifies the plume. These results illustrate our approach's generalization, which will be key to providing a solution with a high temporal resolution.

\end{document}